\documentclass[prd,aps,showpacs,epsf,amsfonts,floats,
amssymb,amsmath,nofootinbib]{revtex4}

\usepackage{color}
\usepackage{graphicx}
\usepackage{amssymb}
\usepackage{hyperref}

\DeclareRobustCommand*\textsubscript[1]{%
  \@textsubscript{\selectfont#1}}
\def\@textsubscript#1{%
  {\m\ensuremath{_{\mbox{\fontsize\sf#1}}}}}

\newcommand{\be}{\begin{equation}}\newcommand{\ee}{\end{equation}}
\newcommand{\bea}{\begin{eqnarray}}\newcommand{\eea}{\end{eqnarray}}
\newcommand{\brr}{\begin{array}}\newcommand{\err}{\end{array}}
\newcommand{\bit}{\begin{itemize}}\newcommand{\eit}{\end{itemize}}
\newcommand{\ben}{\begin{enumerate}}\newcommand{\een}{\end{enumerate}}

\definecolor{darkred}{rgb}{.8,0,0}

\definecolor{darkblue}{rgb}{0,0,.8}

\def\lab{\label}

\def\lf{\left}

\def\pa{\partial}
\def\rar{\rightarrow}
\def\ri{\right}

\def\al{\alpha}

\def\la{\lambda}

\def\1{{_{1}}}\def\2{{_{2}}}

\begin{document}

\title{Collective Molecular Dynamics of  a Floating Water Bridge}

\author{Emilio Del Giudice${}^{a}$, Elmar C. Fuchs${}^{b}$\footnote{Corresponding author. Email: elmar.fuchs@wetsus.nl} and Giuseppe Vitiello${}^{c}$}


\vspace{2mm}



\address{${}^{a}$ Istituto Nazionale di Fisica Nucleare, Sezione di Milano, Milano - 20133 Italy\\and IIB, Neuss, Germany\\
${}^{b}$ Wetsus, Centre of Excellence for Sustainable Water Technology, Agora 1, 8900 CC Leeuwarden, The Netherlands\\
${}^{c}$ Dipartimento di Matematica e Informatica and INFN,
 Universit\`a di Salerno, Fisciano (SA) - 84084 Italy
}

\begin{abstract}
When a high voltage is applied to pure water filling two beakers kept close to each other, a connection forms spontaneously, giving the impression of a floating water bridge. This
phenomenon is of special interest, since it comprises a number of phenomena currently tackled
in modern water science. The formation and the main properties of this floating water bridge are analyzed in the conceptual framework of quantum electrodynamics. The necessary conditions for the formation are investigated as well as the time evolution of the dynamics. The predictions are found in agreement with the observations.
\end{abstract}



\maketitle

\section{Introduction}

In 1893 Sir William Armstrong placed a cotton thread between two wine glasses filled with chemically pure water. After applying a high voltage, a watery connection formed between the two glasses, and after some time, the cotton thread was pulled into one of the glasses, leaving, for a few seconds, a rope of water suspended between the lips of the two glasses \cite{Armstrong93}. As gimmick from early days of electricity this experiment was handed down through history until the present authors learned about it from W. Uhlig, ETH Z{\"u}rich \cite{Uhlig05}. Although easy to reproduce, this watery connection between the two beakers, which is further referred to as 'floating water bridge' holds a number of interesting static and dynamic phenomena \cite{Fuchs07,Fuchs08,Fuchs09,Nishiumi09,Woisetschläger09,Fuchs10,Fuchs10b}.
At macroscopic scale several of these phenomena can be explained by modern electrohydrodynamics, analyzing the motion of fluids in electric fields (see, e.g., the Maxwell pressure tensor considerations by Widom et al. \cite{Widom09}, or the book of Castellanos \cite{Castellanos98}), while on the molecular scale water can be described by quantum mechanics (e.g. \cite{Mrazek06,Jorgensen05}). The gap at mesoscopic scale is bridged by a number of theories including quantum mechanical entanglement and coherent structures in water, theories which are currently discussed (e.g. \cite{DelGiudice06,Head-Gordon06,Stanley05,Dreismann97,Arani95}). Recently, {  a} 2D neutron scattering study indicated a low-level long-range molecular ordering within a $D_{2}O$ bridge \cite{Fuchs10}. Detailed optical measurements \cite{Woisetschläger09} suggested the existence of a mesoscopic bubble network within the water bridge. The properties of water at these scales have drawn special attention due to their suggested importance to human physiology \cite{Pollack01}.
In this paper, we consider the interaction of an applied high voltage potential with the water molecules by exploring the suitability of a quantum field theory  approach, in particular a quantum electrodynamics (QED) approach  to the structure of liquid water proposed in refs. \cite{PRL88,PRA2006,Arani95}.

The paper is organized as follows. In section \ref{SecExp} the experimental set-up and the measurement methods are described, in section \ref{SecII} the appearance of coherent structures in liquid water as an outcome of QED is discussed. In Section \ref{SecIII} we analyze in the same framework the formation of a mesoscopic/macroscopic vortex in liquid water as a consequence of the application of high voltage. Sections \ref{SecIV} and \ref{SecV} are devoted to the analysis of the process of formation of the water bridge including a comparison between theory and experiment, and to the understanding of some of its properties, respectively. Some conclusions are finally drawn in Section \ref{SecVI}.

\vspace{0.5cm}

\section{Experimental} \label{SecExp}

For the high speed imaging experiments performed, flat platinum electrodes were submerged in the center of the beakers, one set to ground potential (cathode), the other on high voltage, up to $25 ~kV$ dc (anode). For the thermographic measurements, cylindric silver electrodes were used. The beakers were filled with deionized $H_{2}O$ (`milli-Q' water, conductivity $\leq 1µS/cm$). A Phywe high--voltage (HV) power supply ('Hochspannungs--Netzger{\"a}t $25 ~kV$', Order No 13671.93) was used with a $42 ~nF$ ceramic capacitor connected in parallel to the electrodes. The voltage was measured by a potential divider of $500M \Omega /500 ~k \Omega$ to ground level. Since the voltage generator provided a limited current output, the electric current was stable at $0.5 ~mA$, while the voltage continuously adapted. The images in Fig. 1 and Fig. 2 were recorded with a FLIR 620 thermographic camera  (FLIR Systems, Boston, MA, USA). The images in Fig. 3 and 4 were recorded with a Photron SA-1 high speed camera (Photron Ldt, Bucks, United Kingdom). Fig. 3 was recorded with $2000 ~fps, ~1/5000 ~s$ exposure time, after $0 ~ms$ (a, reference point), $198 ~ms$(b) $268 ~ms$(c) $306 ~ms$(d) $536 ~ms$(e) $551 ~ms$(f) $626 ~ms$(g) and $1020 ~ms$(h). Fig. 4 was recorded with $1000 ~fps$, $1/1000 ~s$ exposure time, after $0 ~ms$ (a, reference point), $510 ~ms$(b) $520 ~ms$(c) $570 ~ms$(d) and $630 ~ms$(e). In all cases, the high voltage was manually increased from $0$ to $15 ~kV$ (+) DC, and the images were taken at the moment of bridge or vortex creation, respectively, which was between $7$ and $11 ~kV$.

\section{Coherent structures in liquid water} \label{SecII}

In this section we focus our attention
on the gauge invariant properties of a system, like the water bridge, which exhibits complex dynamics. A key motivation for this approach
arises from the fact that gauge invariance is the basic requirement to be satisfied when dealing with systems where charge density and electric polarization density play a relevant role.

It is apparent that the discussion of the gauge invariant properties can be done only within a field description of the structure and dynamics of water. As a matter of fact, many models introduced so far to describe water are based on molecular dynamics which is an approximation that does not consider  the field features of water (for reviews see e.g. \cite{Teixeira,Cabane05}). A conceptual step into this direction was the experimental proof for quantum entanglement in liquid water at room temperature: Chatzidimitriou--Dreismann et al. (1995) did Raman light-scattering experiments \cite{Dreismann95} on liquid $H_{2}O$-$D_{2}O$ mixtures which provided experimental evidence for the quantum entanglement of the
ion $OH^{-}$ (and
$OD^{-}$) vibrational states, and in 1997 a first experimental proof of nuclear quantum entanglement in liquid water \cite{Dreismann97} was published, again, by Chatzidimitriou--Dreismann et al. by the means of inelastic neutron scattering. The interpretation is disputed \cite{Torii00,Dreismann00}. Another approach where water is considered a `hot quantum liquid' was proposed in 2006 \cite{Yoon06}.

In the frame of the theory of liquids the model of liquid helium proposed by Landau \cite{Landau} is appealing. Within this model the liquid appears as made up of two phases, one coherent (having components oscillating in phase), the other non-coherent (having independent components as in a gas). There is no sharp space separation between the two phases since a continuous crossover of molecules occurs between them. This dynamical feature makes the experimental detection of the two phase structure a delicate task indeed. As a matter of fact, an experimental probe having a resolution time longer than the typical period of the particle oscillation between the two phases produces a picture which is an average of the conformations assumed by the system during this time and produces the appearance of an homogeneous liquid \cite{Hendricks74,Bosio81}. On the contrary, the two phase structure would be completely revealed by an instantaneous measurement only. In a realistic situation, an observation lasting a time short enough could give evidence of the chunks of the coherent phase which succeeded to remain coherent during the whole time of the measurement. This kind of observation would give some evidence of the existence of a two phase structure, but would be not enough to give the full instantaneous  extension of the coherent region.

Recently, two articles \cite{PNAS,Israeliani} in favor of the proposed model appeared. In ref. \cite{PNAS} 
evidence of two phases of water having different densities and orderings is presented. Ref. \cite{Israeliani} discusses a comprehensive account of the experimental data supporting the existence in liquid water of aggregates quite larger than those accountable in terms of customary electrostatic theories.
In the frame of QED \cite{PRL88,PRA2006,Arani95} a description of liquid water exhibiting two interspersed phases in agreement with these last experimental findings has been worked out. The two phases are:

$i)$ a coherent phase made up of extended regions, the so called "coherence domains" (from now on referred to as CDs) where all water molecules oscillate in phase between two  configurations.

$ii)$ a non-coherent phase made up of independent molecules trapped in the interstices among the CDs.

The coherent oscillations of the molecules belonging to the first phase is maintained by the electromagnetic (e.m.) field self-produced and self-trapped within the CD, and occur between two definite molecule states. So far two such processes have been identified:

$a)$ in a process analyzed in Ref. \cite{PRL88,PRA2006} the oscillations occur between two rotational levels of the water molecule producing correlations ranging up to several hundreds of microns and giving rise to a common dipole orientation of the molecule electric dipoles which, as a consequence of the rotational invariance of the aqueous system, give rise to a zero net polarization. However, when the rotational symmetry is broken by an externally applied polarization field like near an hydrophilic membrane or near a polar molecule, a permanent polarization develops whose range depends on the amplitude of the external polarization field and hence on the physical state of the surface. The application of a voltage would obviously increase this kind of correlation.

$b)$ in a process analyzed in Ref. \cite{Arani95} the oscillations occur between the ground electronic state of the molecule and the excited $5d$ state at $E_{exc} = 12.06~eV$, just below the ionization
threshold at $12.60~eV$. Each CD has a size given by the wavelength of the resonating e.m. mode $\lambda = \frac{hc}{E_{exc}} =
0.1~\mu m$. The frequency of the common { oscillation of molecules and
field is $0.26~eV$ (in energy units) at $T = 0 ~ K$, much lower than the frequency, $12.06~eV$, of the
free field.  This dramatic softening of the e.m. mode is just the consequence of the nonlinear dynamics occurring in the process
\cite{PRL88,Arani95,PRA2006}.

Non aqueous molecules cannot participate in the resonant dynamics, so they are excluded from the CDs. In particular atmospheric gases are expelled from the CD volume and give rise to a nano- and/or micro bubbles adjacent to the CD. When the dynamics cause a decoherence of a CD, such bubbles, too, would disappear, since their components are able to dissolve again in the non coherent fraction of water.

The amount of the coherent fraction in the liquid is decreasing
with temperature. At room temperature the two fractions
are approximately equal \cite{Arani95}.
In the bulk water, molecules are subjected to two
opposite dynamics: the electrodynamical attraction produced by
coherence and the disruptive effect of the thermal collisions, so
that, whereas in the average the relative fractions are time
independent, at a local, microscopic level each molecule is
oscillating between the coherent and the non-coherent regime.
The coherent structure is thus a flickering one, so that an
experiment having a duration longer than the life time of a CD
probes water as an homogeneous medium.

In the coherent state in the fundamental configuration, whose
weight is $0.87$, all the electrons are tightly bound, whereas in
the excited configuration, whose weight is $0.13$, there is one
quasi-free electron. Consequently a CD contains a reservoir of $0.13
\times n$ quasi-free electrons. At room temperature $n$ is
about $6 \times 10^{6}$. In ref. \cite{Boston} it has been shown
that this reservoir can be excited producing cold vortices of
quasi-free electrons confined in the CD. The energy
spectrum of these vortices  can be estimated following the mathematical scheme outlined in ref. \cite{Boston}. Similarly, it can be seen that the lowest lying excited state has a rotational frequency of a few $kHz$ and the spacing of the levels has the same order of magnitude.
The life-time of these vortices can be extremely long because coherence prevents random (thermal) fluctuations  and because the conservation of the topological charge prevents the decay of the vortex in a topologically trivial state.

We finally remark that in a coherent region the pure gauge nature of the
e.m. potential field
\be \label{empot}
A_{\mu} \approx \pa_{\mu} \la ~,
\ee
where $\la$ is the gauge transformation function, implies that the applied electric potential $V$ is proportional to the time derivative of the phase $\phi$ (see below): $V \propto \frac{d \phi}{dt}$.

The two above dynamics occur simultaneously and produce a non-trivial  interplay that gives rise to a number of phenomenological situations:

$1)$ {\it Bulk water in normal conditions}. In this case the dynamics $a)$ {  are} phenomenologically irrelevant due to the rotational symmetry of the system. The dynamics $b)$ only {  are} at work, producing instant CD structures as large as $0.1$ $\mu m$, which are exposed to the disruption of the thermal collisions giving rise to a Landau-like situation where water appears homogeneous in experiments having a resolution time long enough and exhibiting deviations from homogeneity at smaller resolution times. Moreover, the flickering nature of the coherent structure prevents the appearance of the long time features of the coherent dynamics. In the case of the bulk water the experimental check of the theory is the correct prediction of the thermodynamic processes, which do not depend on the space distribution of the coherent molecules, but on their total number only. The flickering space structure of CDs implies that the corresponding ensemble of microbubbles described above should be a flickering one too, as
found by the experimental observations.

$2)$ {\it Interfacial water}. In this case, dynamics $a)$ {  are} at work and {  their} interplay with dynamics $b)$ gives rise through nonlinear dynamics to a stabilization of the coherent structure which is much more protected against thermal fluctuations. The equilibrium between the two phases is shifted toward coherence, since an extra energy gap is added, molecules are kept aligned by the total polarization field produced by the dynamics $a)$ that compels the radiative dipoles produced by the dynamics $b)$
to stay aligned. The smaller CDs ($0.1$ $\mu m$) of dynamics $b)$
are tuned together by the much more extended coherence produced by dynamics $a)$, so that the global coherent region, thanks to the polarization field produced by the surface gets widened up to several hundreds microns, which is the span of the CD dynamics $a)$. A 
confirmation of this QED prediction is provided by the experimental findings of the group led by G. H. Pollack \cite{PollackZheng}, which confirm results obtained more than sixty years earlier \cite{Henniker}. They show that layers of ``anomalous'' water (EZ water) as thick as $500$ $\mu m$ appear on hydrophilic surfaces. The observed anomalies include the exclusion of solutes, an highly reducing power (corresponding to a negative redox potential of several hundreds of millivolts), and widely different optical and electrical properties. These anomalies are compatible with those expected from the coherence domains of dynamics $b)$ \cite{Arani95}. A much more detailed discussion of this important point will be given elsewhere.

$3)$ {\it Bulk water in special conditions}. Ref. \cite{DelGiudTedeschi} describes the possibility of the onset of a coherence among CDs induced by the tuning of the phases of the oscillations of the CDs, which in normal water are not correlated. This tuning of the different CDs can also be induced by the application of an external e.m. field. A recent 2D neutron scattering study indicated a preferred molecular orientation within a heavy water bridge \cite{Fuchs10}. This observation can be interpreted in accordance with the dynamics $a)$. Moreover, this prediction could account for the experimental observation of a so called ``Neowater'' produced by an Israeli  group \cite{Katzir} and compatible with similar results of Russian \cite{Korotkov} and Ukrainian researchers \cite{Andriewski}. This important point, too, will be discussed at length elsewhere.

The stabilization of the array of water CDs implies the stabilization of the corresponding ensemble of microbubbles, which  therefore form a stable and ordered array. This result has been reported in Ref. \cite{Katzir}. Katzir et al. connect 
the ordered nature of the neowater structure with the appearance of the ordered network of microbubbles; they report as a typical size of the single microbubble a value comparable to the CD size. Thus QED provides a rationale for this surprising phenomenon.

The cases $2)$ and $3)$ are however different. In the case $2)$ the superposition of the two coherent dynamics gives rise to a continuous coherent region, which doesn't contain non coherent zones and therefore, there are no bubbles either. In the case $3)$ there is a coherent ensemble of CDs, which allows the presence of interstices and microbubbles.

In this paper we concentrate on describing
what happens in the particular case of the floating water bridge. This discussion will be done in the next Section.

\vspace{0.5cm}

\section{Formation of a mesoscopic/macroscopic vortex} \label{SecIII}

{ The formation of the floating water bridge is triggered by the application of a high voltage $V$. Let us discuss which are the consequences of such a physical operation on a coherent structure such as the one described in the previous Section. We split the e.m. potential $A_{\mu}$ into the electric and magnetic part. In $cgs$ units we have
\be \label{V}
V = - \frac{\hbar}{e} \frac{d \phi}{dt}~,
\ee
\be \label{Va}
{\bf A} = \frac{ \hbar c}{e}~ {\bf grad ~\phi}~,
\ee
where $e$ denotes the electron charge. According to Eq. (\ref{V}) the application of a voltage implies a strong variation of the phase $\phi$ which adds up to the original phase of the unperturbed CDs. Should the applied potential be high, the voltage generated phase would be dominant with respect to the original phase, which might be regarded as a small perturbation of the total phase. The new phase spans over a macroscopic region and is thus space-correlating all the phases of the CDs enclosed in the macroscopic region. A coherence among the CDs emerges. Moreover, in this new macroscopic coherent region a definite non-vanishing gradient of the phase has appeared, that in turn, according to Eq. (\ref{Va}), produces a non-vanishing magnetic potential. The presence of a magnetic field depends on the rotational or irrotational character of the geometry of the problem. One can show \cite{NuclPhys1975,NuclPhys1975B} that
consistency with the gauge invariance requires that
\be\lab{vs9f} \phi(x) \rar \phi(x) - e \al f(x) ~, \ee
where $\al$ is a constant depending on the wave renormalization constant and the transformation function $f(x)$ is a solution of the equation $\pa^2 f(x) =0$. We use $x \equiv ({\bf x}, t)$. The macroscopic current $j_{\mu,cl}$ is given by \cite{NuclPhys1975,NuclPhys1975B}
\be\lab{g3.vs21} j_{\mu,cl}(x)
=\,
 m_V^2 \lf( a_\mu(x) - \frac{1}{e} \pa_\mu f(x) \ri)~,
\ee
where $a_\mu(x)$ is the classical e.m. field which has acquired the mass $m_V$ (the Anderson--Higgs-Kibble mechanism \cite{Anderson}). One has $\pa^{\mu}j_{\mu,cl}(x) = 0$.

It is crucial here to stress that the presence of quasi--free electrons in the elementary CDs fully characterizes the dynamical regime of the system.  Since these quasi--free electrons are confined within the CDs, their motion is necessarily a closed one, which implies that a magnetic field is thus generated. This fact, together with the non--trivial topology of the beaker system due to the presence of the (charged) electrode in the beaker center, produces a macroscopic vortex.  Such a vortex is indeed present in both beakers and can be visualized with a thermographic camera as shown in Fig. 1.

Thus, an extended coherence on a scale much larger than the original $0.1$ $\mu m$ is the consequence of the application of a high voltage, also in the absence of an electric current, as is the case in the short time just after engaging the voltage and before the appearance of the bridge.
This effect can be seen as a variant of the well known Bohm--Aharonov effect \cite{Sakurai}.

Such a phenomenology is described in formal terms by the fact that the function $f(x)$ may indeed carry a topological singularity describing the occurrence of a vortex and given by
\be \lab{fx} f(x) = \arctan \left(\frac{x_{2}}{x_{1}}\right)~.
\ee
Eq.~(\ref{fx}) shows that the phase is undefined on the line
$r = 0$, with $r^{2} = x^{2}_{1} + x^{2}_{2}$, consistently with the
phase indeterminacy at the electrode site due to the specific system
geometry.

When $f(x)$ carries a (vortex) topological singularity, it means that it is
is path-dependent (not single-valued). When  $f(x)$ is a regular function, i.e. it does not carry a topological singularity, then the current $j_{\mu,cl}$ vanishes \cite{NuclPhys1975,NuclPhys1975B}, which in turn implies zero e.m. field $F_{\mu\nu} = \pa_{\mu} a_{\nu} - \pa_{\nu} a_{\mu}$.

In conclusion, in the dynamical scenario depicted above, the
action of turning on the high potential compels the CDs
to join the observed giant vortex  whose core position is
determined by the electrode position. The closer the CDs are to the
vortex core (the electrode), the stronger is the action
pushing them to join together. Since the applied potential is a decreasing function of the distance from the electrode, peripheral CDs (those nearest to the beaker wall) have better chance to preserve their individuality
(although {  they are} in a non--equilibrium regime due to the criticality of the dynamics
going on). It should be pointed out that the presence of
the other electrode in the grounded beaker and the consequent existence
of a preferred radial direction on the joining line between the
two electrodes break the cylindrical symmetry around each one of
the electrodes.

\begin{figure}[htbp]
\centering \resizebox{18cm}{!}
{\includegraphics{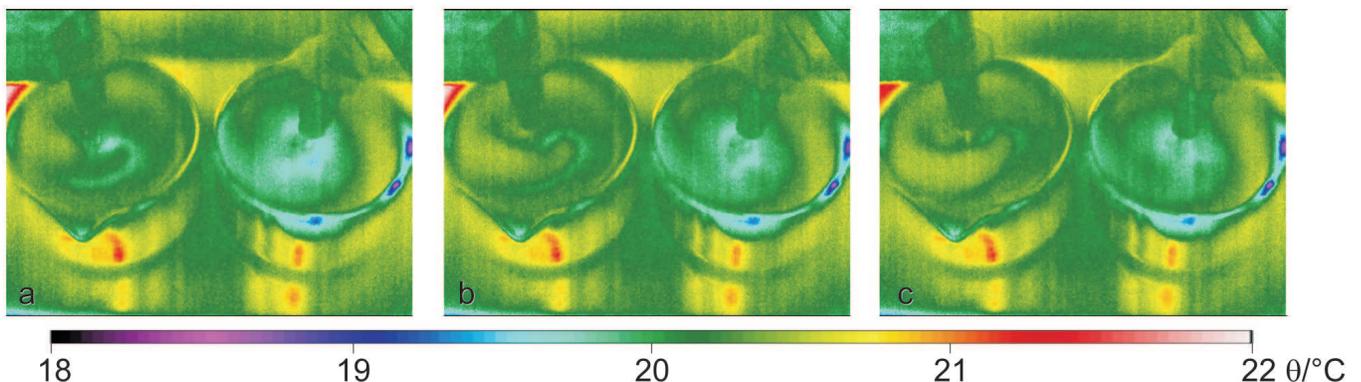}}
\caption{Thermographic vizualisation of the macroscopic Vortex in the beakers under application of high voltage before the water bridge is formed. The time interval between the images is 5 sec, the temperature scale is calibrated to the emissivity of water (0.96).}
\label{fig1}
\end{figure}

The radial velocity of the slightly ($\approx1K$) colder vortices is $\approx 1^{\circ}/s -3^{\circ}/s$ and could be observed independently of the potential applied. The rotational direction was counter-clockwise in both beakers. In Fig. 2, the symmetry break due to the influence of the second beaker is visualized by thermography. Fig. 2a again shows the slight cooling of the vortices. After a certain threshold (between $9$ and $11 ~kV$), a cooling along the joining line of the electrodes as shown in Fig. 2b appears ($2-3 ~K$), and the rotation stops. The cooling is then followed by electric discharges heating the beaker walls (Fig. 2c) and finally leads to the water bridge formation (Fig. 2d). The dark spots on the anode in Fig. 2c and d are water droplets which were ejected from the beakers during the process.

\begin{figure}[htbp]
\centering \resizebox{18cm}{!}
{\includegraphics{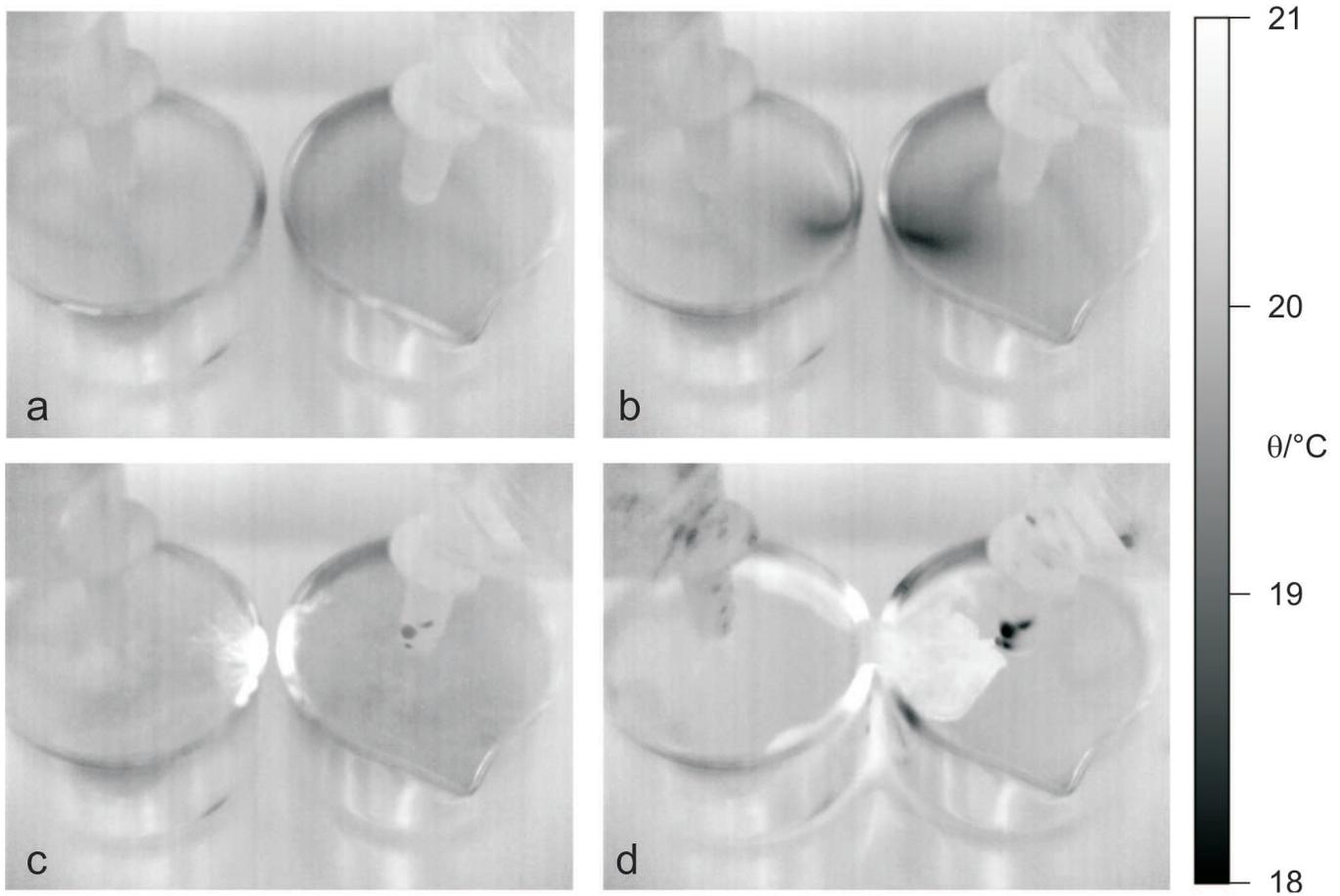}}
\caption{Thermographic visualization of the bridge formation mechanism. First, the macroscopic vortices appear (a), then the water cools down at the joining line of the two electrodes (b). With the first sparks (c), the water heats up, and finally forms a water bridge (d). The time interval between the images is $\approx 1 ~s$, the temperature scale is calibrated to the emissivity of water ($0.96$). The dark spots on the electrods in (c) and (d) are water droplets which were ejected during the process.}
\end{figure}

The vortices occur independently of the value of the
applied potential.
The potential acts
as a trigger inducing the phase transition. The inner dynamics of the water then {  control} the system
evolution and its vorticity. This explains the observed independence of the turbulent patterns of the strength of the applied
potential.

Finally, we note that the above description which starts from the analysis of the microscopic dynamics is consistent with the results obtainable by use of the classical electro--hydro--dynamics (EHD) field equations, originally proposed by Melcher and Taylor \cite{Melcher69} and completed by Saville \cite{Saville97}, describing the effects of a high voltage applied onto a ''leaky dielectric''.

\vspace{0.5cm}

\section{Formation of the water bridge}\label{SecIV}

It is well known \cite{Anderson} that within a coherent region a magnetic field should vanish (Meissner effect), provided that the size of the region exceeds a threshold (London penetration length). Indeed, the magnetic field penetration in the coherent region decays exponentially as described by the London equation
\be
\nabla ^{2} {\bf H} = \frac{1}{\la^{2}_L} {\bf H}
\ee
where $\la_L$ is the London penetration length (see, e.g., Ref. \cite{Feynman}). This property is the consequence of the regularity of the phase in the coherent region (far from the boundaries), which produces the vanishing of the magnetic field (cf. Eq.~(\ref{g3.vs21}) and the comments following Eq.~(\ref{fx})). Near the boundaries the phase acquires a singular behavior due to dishomogeneities and this produces a non vanishing magnetic field (cf. Eq.~(\ref{fx})).  In the case of liquid water $\la_L$ has a value much larger than the size $0.1~\mu m$ of the water CD, in agreement with the well known fact that magnetic fields are not expelled from normal water.

However, when the peculiar dynamics outlined in the previous section {  are} at work, the coherent region could become so far extended that its size  overcomes the London penetration length. In this case a phenomenon of levitation analogous to the one observed with superconductors might occur. Actually in the presence of  the Meissner effect a gradient of the magnetic inductance $\mu$ appears since $\mu = 0$ in the coherent region and is about one outside. A magnetic levitating force
${\bf F}_{lH} = -H^{2} {\bf  grad}~ \mu$ appears, provided that $H^{2}$ is inhomogeneous. At the air water interface the presence of the macroscopic vortex described above produces an high value of $H^{2}$ below the water surfaces, whereas in the atmosphere above the surface $H^{2}$ is much lower and corresponds to the ambient magnetic field. Consequently, a net upward force develops that raises the CD up. Since the extended coherent region is the one close to the beaker wall, magnetic levitation would occur along the wall of the vessel where the size of the coherent regions is larger.  These magnetic forces
would occur only in the vessel with the high positive potential where the vortex can develop, whereas will be absent in the grounded beaker.

Moreover, also an electric levitation force
\be \lab{fe}
{\bf F}_{le} = - E^{2} {\bf  grad}~ \epsilon
\ee
may appear, since the dielectric constant $\epsilon$ is much larger in the CDs than in the non coherent region. As a matter of fact, if we model the non coherent region as an ensemble of independent electric dipoles, the Fr\"ohlich formula would gives us at room temperature $\epsilon = 15$, whereas the dielectric constant of the CDs could be estimated to be 160 \cite{unpublishedEDGPrepar}. Since at room temperature the two fractions are almost equal the two above estimates result in an average dielectric constant $\epsilon_{obs}$ of $\epsilon_{obs} = \frac{160}{2} + \frac{15}{2} = 87.5$ {  which is} in good agreement with the experiment.

Along the line joining the two electrodes the electric field is very strong and reduced below the water surface by a factor $\frac{1}{\epsilon}$, whereas {  it} is at full strength above the air water surface. A net upward force is thus generated according to Eq.~(\ref{fe}).

In conclusion, we conclude that the application of a high electric voltage, through the complex dynamics outlined above, gives rise to two levitations, electric and magnetic along the wall of the vessel with the positive electrode, and to one levitation, electric only, in the grounded vessel. This prediction is in agreement with the observation of a larger probability of  water column formation of in the HV vessel than in the grounded one
as shown by high-speed visualization of the examples given in Fig. 3 and Fig. 4. For both figures, fringe projection technique is used to contour the bridge and thus enhance its visibility (for details see \cite{Woisetschläger09}).

\begin{figure}[htbp]
\centering \resizebox{18cm}{!}
{\includegraphics{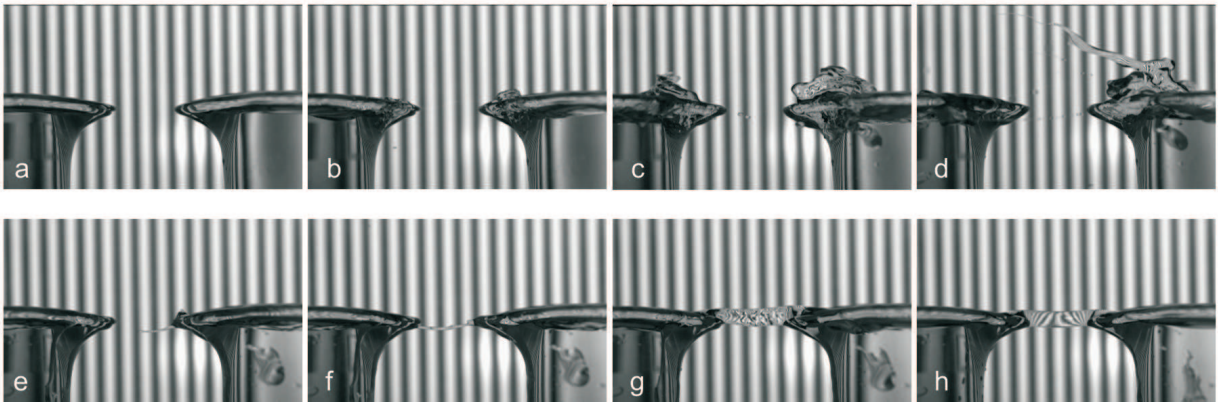}}
\caption{High speed vizualisation of the water bridge formation with glass beakers with fringe projection. (a) shows the beakers before the voltag is applied, (b) -- (d) show the levitation of the water in both beakers leading to droplet formation, (e)--(f) show the ejection of a jet from the HV beaker leading to the bridge formation (g), which is stabilized in (h). The water on the HV side (right) is levitated stronger than on the grounded side (left) due to the fact that the water is lifted both electrically and magnetically there, whereas on the grounded side, the levitation is due to electric forces only.  }
\end{figure}

The levitating drops of water, being coherent, are surrounded by the electromagnetic evanescent field, filtering out of the coherent cores. The tail of the evanescent field
spans for a length of the same order of the CD radius, so that it could act as an interaction field among the drops distant as much. This distance is in the order of the droplet radius, namely some microns.  Therefrom the possibility of the formation of a string of interacting water droplets emerges that eventually give rise to the water bridge.

\begin{figure}[htbp]
\centering \resizebox{18cm}{!}
{\includegraphics{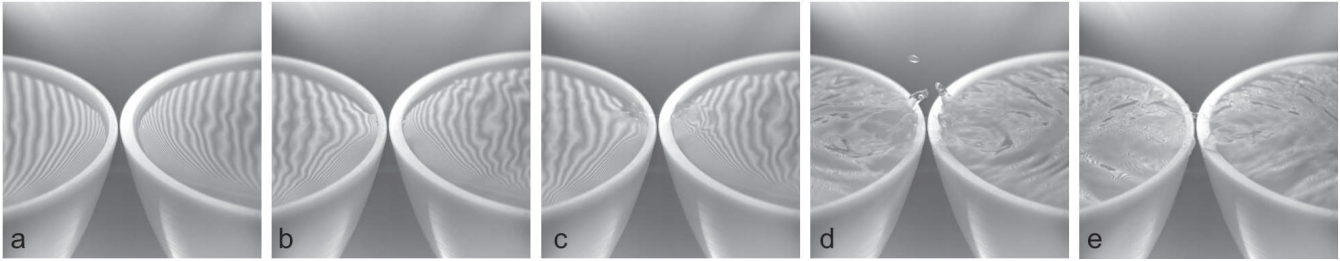}}
\caption{High speed visualization of the water bridge formation with teflon beakers with fringe projection. (a) shows the situation without high voltage, in (b) and (c) the fringe projection shows the rising of the surfaces, in (d) levitation and droplet formation are shown, and in (e) a connection is finally formed.}
\end{figure}

\vspace{0.5cm}

\section{Properties of the water bridge}\label{SecV}

The water bridge thus represents a sequence of interacting coherent droplets extracted from the water vessels. The electron coherence illustrated in Section \ref{SecII} points to the CDs as reservoirs of quasi free electrons. On the CD boundaries a ponderomotive force
\be \lab{fp}
{\bf F}_{p} = - \frac{q^{2}}{2m} ~{\bf  grad} ~|{\bf A}|^{2}
\ee
acts upon any particle having charge $q$ and mass $m$ present there. Eq.~(\ref{fp}) can be easily understood considering that the Hamiltonian of a particle immersed in a vector potential
gives rise to a field energy distribution $U = \frac{q^{2}}{2m} |{\bf A}|^{2}$ which produces in turn the force ${\bf F}_{p} = - {\bf  grad}~ U$.

The ponderomotive force Eq.~(\ref{fp}) pushes  the quasi free electrons outwards with a force many thousand {  times} larger than that acting on the parent molecules. As a consequence, a double layer of charges appears on the boundary of the CDs. The applied strong electric field induces a twofold motion along the bridge. The positive cores of the CDs are pushed along the bridge from the anode to the cathode giving rise to a simultaneous transport of mass, whereas the outer negatively charged layer slides toward the cathode. We should remember that the component droplets of the water bridge have been extracted from a macroscopic vortex so that their motion along the bridge arises from the superposition of the vortex motion and the electrically induced motion. The final result is an helicoidal motion around the axis of the bridge. A sort of traveling vortex. According to the topological considerations developed in Section \ref{SecII}, the axis of the bridge should be the site of a topological singularity, namely no coherence should exist on the axis. Consequently, a negative gradient  of the coherent fraction should be observed when going from the outer surface to the axis of the bridge. This prediction is consistent with observation of a negative gradient of the speed of sound in the same direction \cite{Fuchs10b}. Actually, since the coherent region is very much correlated, the speed of sound should be larger in the coherent region than in the non coherent one.

The peculiar optical properties of the bridge are described in Ref. \cite{Woisetschläger09}, in particular the 
change of the polarization angle of linearly polarized light when passing through the bridge. As suggested in Ref. \cite{Fuchs08}, this change can be connected to both reflection at micro-bubbles under the Brewster angle \cite{Woisetschläger09}, or, {  as shown} in a very recent work, to birefringence due to long-range low-ordering within the bridge \cite{Fuchs10}.
A detailed discussion of this in the framework of QED will be given elsewhere.

Experimental observations \cite{Woisetschläger09} have shown that cooling, achieved by, e.g., the addition of icecubes to the beakers, destabilizes the bridge. This can be easily explained within the proposed model. The coherent fraction increases when temperature decreases as well as the depth of the interstices among them. Consequently the excitation of rotational motion of the CDs becomes more and more difficult because of the drag produced by the interaction with the neighboring CDs through the evanescent fields protruding from inside them. At the freezing point the drag becomes so intense to prevent the formation of vortices.  As explained, the existence of these vortices is crucial for the presented analysis, their reduction explains the destabilization of the phenomenon.

Finally, the  suggested ordered network of microbubbles would be a natural consequence of the presence of a permanent extended coherence as discussed in Section \ref{SecII}.

 \vspace{0.5cm}

\section{Conclusions}\label{SecVI}

Much work is still needed to understand the wealth of the results revealed by the water bridge. However, in this paper we have shown that a consistent framework for understanding this surprising phenomenon can be provided by QED.

Without doubts, water is one of the most common and most studied substances in the world. The properties of water at mesoscopic scale have drawn special attention lately due to their suggested importance to human physiology \cite{Pollack01}. Still, present theories have difficulties explaining more than a few of its properties at once, and no theory so far could satisfyingly explain one lately rediscovered phenomenon, the floating water bridge. The QED approach to this phenomenon provides a possible theoretical background for many of the bridge's features:
\begin{itemize}
	\item The vortex formation upon applying a potential
	\item The occurrence of a cold region prior to the rising of the water
	\item The asymmetric rising of the water in the beakers
	\item The stability of the bridge
	\item The temperature dependence of this stability
	\item The correlation of charge and mass transfer
	\item The formation of micro and nanobubbles and consequently
	\item The findings by optical and neutron scattering
\end{itemize}
Therefore we think that, although unconventional in the field of electro-hydro-dynamics (EHD), quantum field theory can be a powerful tool to bridge the gap between microscopic descriptions and field theories, something that has been conceived to be a quite difficult task \cite{Castellanos98} until now.
Moreover, some of the described effects can be found in other experiments as well, like the macroscopic vortices discussed herein which are, although hitherto unexplained, known in nowadays EHDs, even an application as motor was discussed recently \cite{Sugiyama08}; or the suggested formation of a clear zone next to gel surfaces found by the Pollack group \cite{Pollack01}.
Recently Cabane et al. stated that water has thus far been a \textsl{fantastic ''graveyard'' for theories that are clever but wrong} \cite{Cabane05}.  It may be appropriate here to stress the well known fact that no theory can in principle describe reality in all its aspects. Nevertheless, to the knowledge of the authors, no other theory has been able to describe the floating water bridge as correctly in detail as the quantum field theory. Therefore, it may be prudent to assume that QED is a good choice to describe and understand this special interaction of water with electric fields.

\vspace{0.5cm}

\section{Special Acknowledgements}\label{SecAkn}

The authors would like to express there gratitude to Prof. Jakob Woisetschl{\"a}ger for the performance of the presented experiments.

\section{Acknowledgements}\label{SecNakn}

With great pleasure, the authors wish to thank Profs. Marie-Claire Bellissent-Funel (Laboratoire L\'eon Brillouin, Saclay), Eshel Ben-Jacob (Tel Aviv University), Cees Buisman (Wetsus - Centre of Excellence for Sustainable Water Technology), Friedemann Freund (NASA SETI Institute, Mountain View CA), Karl Gatterer (Graz University of Technology), Franz Heitmeir (Graz University of Technology), Jan C.M. Marijnissen (Delft University of Technology), Laurence Noirez (Laboratoire L\'eon Brillouin, CEA-CNRS/IRAMIS, CEA-Saclay), Gerald H. Pollack (University of Washington), Alan Soper (Rutherford Appleton Laboratories, Oxford), Jos\'e Teixeira (Laboratoire L\'eon Brillouin, CEA-CNRS/IRAMIS, CEA/Saclay), as well as Luewton L.F. Aghostinho, Ingo Leusbrock and Astrid H. Paulitsch (Wetsus, Centre of Excellence for Sustainable Water Technology) for the ongoing discussion on the water bridge phenomenon (in alphabetic order). Partial financial support from University of Salerno and Istituto Nazionale di Fisica Nucleare is also acknowledged.


\end{document}